\documentclass[epj]{svjour}
\usepackage{graphicx}
\usepackage{dcolumn}
\usepackage{bm}
\usepackage{amsmath}
\usepackage{color}
\usepackage[normalem]{ulem}
\usepackage{float}

\newcommand{\bra}[1]{\left\langle#1\right|}
\newcommand{\ket}[1]{\left|#1\right\rangle}

\newcommand{\Tr}{{\rm Tr}}

\newcommand{\CC}{{\rm C}}

\newcommand{\RR}{{\rm R}}
\newcommand{\LL}{{\rm L}}

\newcommand{\GammaL}{{\bm\Gamma}_{\rm L}}

\newcommand{\GammaR}{{\bm\Gamma}_{\rm R}}

\newcommand{\Hxc}{{\rm Hxc}}
\newcommand{\xc}{{\rm xc}}

\begin{document}

%\title{Steady state density functional theory for asymmetrically coupled leads}

\title{Exchange-correlation functionals of i-DFT for asymmetrically coupled leads}
\author{Stefan Kurth\inst{1,2,3} \and David Jacob\inst{1,2}}

\institute{Nano-Bio Spectroscopy Group and European Theoretical
  Spectroscopy Facility (ETSF), Dpto. de F\'isica de Materiales,
  Universidad del Pa\'is Vasco UPV/EHU, Avenida Tolosa 72, E-20018 San
  Sebasti\'an, Spain \and
  IKERBASQUE, Basque Foundation for Science, Mar\'ia D\'iaz de Haro 3,
  E-48013 Bilbao, Spain \and
  Donostia International Physics Center (DIPC), Paseo Manuel de
  Lardizabal 4, E-20018 San Sebasti\'an, Spain}

\date{\today} 

\abstract{
  A recently proposed density functional approach for steady-state transport
  through nanoscale systems (called i-DFT) is used to investigate junctions
  which are asymmetrically coupled to the leads and biased with asymmetric
  voltage drops. In the latter case, the system can simply be transformed
  to a physically equivalent one with symmetric voltage drop by a total
  energy shift of the entire system. For the former case, known exchange
  correlation gate and bias functionals have to be generalized 
  to take into account the asymmetric coupling to the leads.
  We show how differential conductance spectra of the constant interaction
  model evolve with increasing asymmetry of both voltage drops and coupling to
  the leads.
}

\maketitle

\section{Introduction}

The measurement of electronic transport through nanoscale devices
provides an important means for probing the electronic and magnetic
structure and related properties of the system.
For example, the measurement of the zero-bias conductance
of atomic-scale metallic nanocontacts formed in break-junction experiments
unveiled conductance quantization and the formation of monoatomic chains
in these systems~\cite{Agrait:PR:2003}.
Conductance spectroscopy of quantum dots coupled to conducting electrodes
demonstrated Kondo effect and Coulomb blockade phenomena~\cite{Goldhaber-Gordon:Nature:1998,Park:Nature:2002,Grabert:book:1992,Champagne:NL:2005}.
Using scanning tunneling microscope spectroscopy (STS), the Kondo effect
and spin excitations of magnetic adatoms and molecules on conducting
substrates can be measured~\cite{Madhavan:Science:1998,Li:PRL:1998,Knorr:PRL:2002,Zhao:Science:2005,Iancu:NL:2006,Hirjibehedin:Science:2007,Parks:Science:2010,Oberg:NNano:2014,Karan:NL:2018}.
Coupling of the electronic degrees of freedom to the nuclear motion even
allows to determine phonon band structures of metals or vibronic 
excitations of molecules by inelastic electron transport spectroscopy (IETS)~\cite{Stipe:Science:1998,Hahn:PRL:2000}.

The theoretical description of electronic transport in nanoscale systems
either involves sophisticated and often computationally demanding
many-body treatments of relatively simple model Hamiltonians, or ab initio
approaches based on density functional theory (DFT) \cite{Lang:95} which
allows to treat realistic systems of substantial sizes (up to thousands of atoms
depending on the implementation).
The description of (steady-state) transport through a nanoscale
system within the framework of DFT goes back to
a seminal paper by Lang \cite{Lang:95}. In this work, following ideas of
Landauer \cite{Landauer:57} and B\"uttiker \cite{Buettiker:86}, steady-state
transport is treated as a scattering problem of effectively non-interacting
electrons where the local Kohn-Sham (KS) potential in the nanoscale region
C is treated as the scattering potential. The resulting scheme, often
called LB+DFT, today is basically the method of choice for the {\em ab initio}
description of electronic transport.
Conceptually, however, there is a
problem: since DFT is a {\em ground state} or {\em equilibrium} theory, there
is no guarantee that the LB+DFT formalism yields the correct
{\em non-equilibrium} density and current, even if the exact KS potential
is used.
Importantly, for this reason many-body phenomena such as the Kondo effect
or Coulomb blockade are not described properly in the \emph{finite bias}
transport characteristics of the system within the LB+DFT
approach~\cite{StefanucciKurth:15}.\footnote{
  It should be noted though that the zero-bias conductance and density
  at zero temperature can in fact be correctly described for strongly correlated systems within LB+DFT
  provided the exact functional or a good approximation to it is
  employed~\cite{Stefanucci:PRL:2011,Bergfield:PRL:2012,Troester:PRB:2012}.}

A combination of the DFT based transport approach with more
sophisticated many-body treatments of model Hamiltonians, allows to incorporate
strong electronic correlations originating from a relatively small subspace (e.g. the
$d$-shell of a transition metal atom), into the description of electronic
transport through realistic systems~\cite{Jacob:JPCM:2015,Droghetti:PRB:2017}. 
The drawback of this approach is that, like all DFT++ approaches, it is hampered
by the infamous double-count\-ing problem and the determination of the
effective interaction strength of the strongly correlated subspace(s),
thus limiting the predictivity of the approach.

It is thus desirable to have an approach that treats the whole nanoscale
system on the same footing, while at the same time being computationally
affordable, in order to be able to treat realistic systems.
In principle, time-dependent DFT (TDDFT) \cite{RungeGross:84} provides
a proper DFT framework to treat non-equilibrium situations such as
transport. Indeed it has been shown that in the steady state which develops
in the long-time limit after switch-on of a DC bias, TDDFT in principle leads
to corrections to the LB+DFT formalism \cite{sa-2.2004,ewk.2004,kbe.2006,skgr.2007}. 

In the present work we will use yet another DFT framework for steady-state
transport, called i-DFT, which has been proposed only recently
\cite{StefanucciKurth:15}.
This approach is comparable to the LB+DFT formalism in computational
effort. However, since i-DFT is an in principle exact approach for the 
out-of-equilibrium steady state, it is capable of describing
many-body effects in transport provided
good approximations to the i-DFT functionals are available.
In fact, already in Ref.~\cite{StefanucciKurth:15} a functional for the
so-called constant interaction model (CIM) has been developed which correctly
describes the Coulomb blockade at arbitrary bias. This approximation
has later been augmented to also include Kondo physics, first for the
the single-impurity Anderson model (SIAM) \cite{Kurth:PRB:2016} and later
for the CIM with an arbitrary number of levels \cite{Kurth:JPCM:2017}.
All the functionals mentioned so far were constructed for the 
case of symmetric coupling to the leads. Very recently,
an approximation for the CIM at extremely asymmetric coupling has
also been developed \cite{Jacob:NL:2018}. 

For the decription of realistic transport setups such as a quantum dot coupled
to leads or a molecule probed by an STM, it is necessary to consider arbitrary
asymmetric coupling to the leads in the i-DFT approach. 
Here we construct i-DFT functionals for the CIM at arbitrary asymmetry in the
coupling to the leads and compare their relative performance.

\section{Density functional theory for steady state transport: i-DFT}

\subsection{Transport setup}

\begin{figure}
  \begin{center}
    \includegraphics[width=0.9\linewidth]{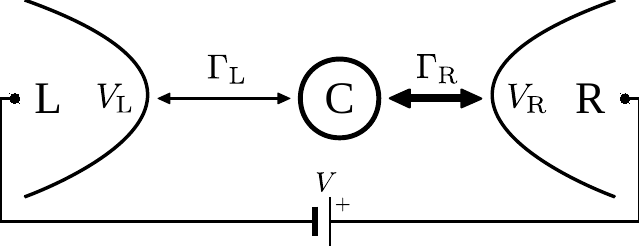}
  \end{center}
  \caption{
    \label{fig:setup}
    Schematic sketch of a nanoscale junction under an applied bias $V$.
    The central region (C) is coupled to two semi-infinite leads L and R.
    The coupling to the left and right leads may differ, i.e. $\GammaL\neq\GammaR$ in general.
    The voltage drop $V$ across the nanoscale junction is defined by the electrochemical
    potentials $V_\LL$ and $V_\RR$ in the two leads, $V=V_\LL-V_\RR$.
  }
\end{figure}

We are interested in the generic situation sketched sche\-matically in
Fig.~\ref{fig:setup} where a central nanoscopic region C is coupled to 
macroscopic left (L) and right (R) leads. The system is driven out of
equilibrium by applying a (DC) bias across region C and we are interested
in the resulting {\em steady-state} current. We emphasize that here we
just assume that the system reaches a steady state after application of the
bias but we do {\em not} address the question how this steady state is
reached (i.e., we are not interested in the time evolution towards the
steady state). The Hamiltonian describing the system coupled to leads
is given by 
\begin{equation}
  \hat{\mathcal{H}} = \hat{\mathcal{H}}_\CC + \hat{\mathcal{H}}_\LL +
  \hat{\mathcal{H}}_\RR + \hat{\mathcal{V}}_{\LL} + \hat{\mathcal{V}}_{\RR}
\end{equation}
where $\hat{\mathcal{H}}_\CC$ describes the nanoscopic region,
$\hat{\mathcal{H}}_\alpha$ describes lead $\alpha$ ($\alpha \in \{L,R\}$)
and $\hat{\mathcal{V}}_{\alpha}$ is the coupling between lead $\alpha$ and
region C. For simplicity, the two leads are considered as non-interacting,
i.e., $\hat{\mathcal{H}}_\alpha=\sum_{q\sigma} \epsilon^\alpha_{q}
\hat{c}_{\alpha,q\sigma}^\dagger \hat{c}_{\alpha,q\sigma}$. In contrast, in the central
region we allow for a general electron-electron interaction such that the
Hamiltonian $\hat{\mathcal{H}}_\CC$ takes the form 
\begin{equation}
  \label{eq:H_C}
  \hat{\mathcal{H}}_\CC = \sum_{i,j,\sigma} h_{ij}^0 \, \hat{d}^\dagger_{i\sigma}
  \hat{d}_{j\sigma} + \frac{1}{2} \sum_{{i,j,k,l}\atop{\sigma,\sigma^\prime}} U_{ijkl} \,
  \hat{d}^\dagger_{i\sigma} \hat{d}^\dagger_{j\sigma^\prime} \hat{d}_{l\sigma^\prime}
  \hat{d}_{k\sigma}
\end{equation}
where $\bm{h}^0=\bm{t}+\bm{v}$ is the one-body part of the Hamiltonian
(in matrix notation) comprising the kinetic energy $\bm{t}$ and external
potential (gate) $\bm{v}$, and $U_{ijkl}$ are the matrix elements of the
electron-electron interaction. The coupling between the central region and
the two leads L and R is described by
\begin{equation}
  \hat{\mathcal{V}}_{\alpha} = \sum_{i,q,\sigma} V^\alpha_{q,i} \,
  \hat{c}^\dagger_{\alpha,q\sigma} \hat{d}_{i\sigma} + {\rm h.c.}
\end{equation}
Integrating out the degrees of freedom of lead $\alpha$ yields the corresponding
embedding self-energy
\begin{equation}
  \label{eq:SE_embed}
  \bm{\Sigma}_\alpha(\omega) = \bm{V}_\alpha^\dagger \frac{1}{\omega - \bm{H}_\alpha} \bm{V}_\alpha
\end{equation}
where $\bm{V}^\alpha=\left(  V^\alpha_{q,i} \right)$ and
$\bm{H}_\alpha=\left( \epsilon_q^\alpha \right)$. The anti-hermitian part of
the embedding self-energy yields the so-called coupling matrix,
$\bm\Gamma_\alpha=i(\bm\Sigma_\alpha^\dagger-\bm\Sigma_\alpha)$, which describes
the broadening of the central region due to the coupling to lead $\alpha$.

Finally, a bias voltage $V$ is applied between the two leads, defined by the
difference in their electrochemical potentials, $V=V_\LL-V_\RR$, which
drives the system out of equilibrium and induces a steady current $I$
across the nanoscale junction.

%In the following the
%i-DFT formalism\cite{Stefanucci:NL:2015,Kurth:PRB:2016,Kurth:JPCM:2017} is applied to study
%the steady state density and current through the nanoscale junction.

%\subsection{Kohn-Sham equations for asymmetric coupling to the leads and symmetric bias}

\subsection{Steady state transport with density functional theory and
  foundation of the i-DFT formalism}

%% The description of (steady-state) transport through a nanoscale
%% system within the framework of DFT goes back to
%% a seminal paper by Lang \cite{Lang:95}. In this work, following ideas of
%% Landauer \cite{Landauer:57} and B\"uttiker \cite{Buettiker:86}, steady-state
%% transport is treated as a scattering problem of effectively non-interacting
%% electrons where the local Kohn-Sham (KS) potential in the nanoscale region
%% C is treated as the scattering potential. The resulting scheme, often
%% dubbed LB+DFT, today is basically the method of choice for the {\em ab initio}
%% description of electronic transport. Conceptually, however, there is a
%% problem: since DFT is a {\em ground state} or {\em equilibrium} theory, there
%% is no guarantee that the LB+DFT formalism yields the correct
%% {\em non-equilibrium} density and current, even if the exact KS potential
%% is used. In principle, time-dependent DFT (TDDFT) \cite{RungeGross:84} provides
%% a proper DFT framework to treat non-equilibrium situations such as
%% transport. Indeed it has been shown that in the steady state which develops
%% in the long-time limit after switch-on of a DC bias, TDDFT in principle leads
%% to corrections to the LB+DFT formalism
%% \cite{sa-2.2004,ewk.2004,kbe.2006,skgr.2007}. 
%% In the present work we will use yet another DFT framework for steady-state
%% transport which has been proposed only recently \cite{StefanucciKurth:15}.
%% This framework, called i-DFT, just like LB+DFT is only concerned with
%% the steady-state density and current for a (DC) biased system.

Here we make use of the i-DFT formalism in order to describe the
steady state density and current of an interacting system under a DC bias~\cite{StefanucciKurth:15}.
In i-DFT, one first establishes (under certain conditions~\cite{KurthStefanucci:17})
the existence of a one-to-one map between the gate potential
$v(\bm{r})$ in region C and the bias $V$ symmetrically applied across it
on the one hand and the steady state density $n(\bm{r})$ and current $I$ on the
other hand. In a second step, just as in standard DFT, one maps the interacting
problem onto a ficticious non-interacting one which exactly yields the
density and current of the interacting system. This fictitious non-interacting
KS system now features two potentials, the KS gate potential $v_s(\bm{r})$
(in region C) and the KS bias $V_s$. Overall this establishes a one-to-one
map between the potentials of the original interacting system
$(v(\bm{r}),V)$ and the ones of the KS system $(v_s(\bm{r}),V_s)$, i.e., 
\begin{equation}
  (v(\bm{r}),V) \longleftrightarrow (n(\bm{r}),I) \longleftrightarrow (v_s(\bm{r}),V_s)
\end{equation}
and therefore both $(v(\bm{r}),V)$ and $(v_s(\bm{r}),V_s)$ are functio\-nals
of $n(\bm{r})$ and $I$. The difference between these functionals allows to
define the Hartree exchange correlation (Hxc) gate potential,
$v_\Hxc[n,I](\bm{r})=v_s[n,I](\bm{r})-v[n,I](\bm{r})$, as well as the
exchange correlation (xc) bias, $V_\xc[n,I]=V_s[n,I]-V[n,I]$. In general, of
course, the exact form of these functionals is unknown and one has to resort
to approximations in practice. 

In the i-DFT framework the steady-state density and current of an
interacting system for given fixed external potential $v(\bm{r})$ and
bias $V$ applied symmetrically in left and right leads can be calculated
by solving the following coupled self-consistent KS equations
\cite{StefanucciKurth:15,Kurth:JPCM:2017}
\begin{subequations}
  \begin{eqnarray}
    n_{\rm sym}[v,V](\bm{r}
    &=& 2\sum_{\alpha=\LL,\RR}\int\frac{d\omega}{2\pi}
    f\left(\omega+s_\alpha\frac{V_s}{2}\right) A_{\alpha;v_s}(\bm{r};\omega)
    \nonumber\\
    %%\hspace{1ex}
    \label{eq:dens}
  \end{eqnarray}
  \begin{equation}
    I_{\rm sym}[v,V] =  2\sum_{\alpha=\LL,\RR} s_\alpha \int\frac{d\omega}{2\pi}
    f\left(\omega+s_\alpha\frac{V_s}{2}\right) T_{v_s}(\omega)
    \label{eq:curr}
  \end{equation}
\end{subequations}
%%
%%with the KS gate potential $v_s(\bm{r})=v(\bm{r})+v_{\rm Hxc}(\bm{r})$
%%and the KS bias $V_s=V+V_{\rm xc}$. 
where $A_{\alpha}(\bm{r};v_s;\omega)=\bra{\bm{r}}\bm{A}_{\alpha;v_s}(\omega)\ket{\bm{r}}$
is the spatial representation of
$\bm{A}_{\alpha;v_s}=\bm{G}_{v_s}\bm{\Gamma}_\alpha\bm{G}_{v_s}^{\dagger}$, 
the partial KS spectral function of C
associated with electron injection from electrode $\alpha$. 
$T_{v_s}(\omega)=\Tr[\GammaL\bm{G}_{v_s}^\dagger\GammaR\bm{G}_{v_s}]$ 
is the KS transmission function and 
$\bm{G}_{v_s}=(\omega-\bm{h}_s-\bm\Sigma_\LL-\bm\Sigma_\RR)^{-1}$ is the 
(retarded) non-equilibrium KS Green 
function of the sample region, and $\bm{h}_s = \bm{t} + \bm{v}_s$ the KS 
Hamiltonian in matrix notation. Finally, $f(x)=[1+\exp(\beta x)]^{-1}$ is
the Fermi function at inverse temperature $\beta$, and 
$s_\alpha=-1$ for $\alpha=\LL$ and $s_\alpha=+1$ for $\alpha=\RR$.

\begin{figure}
    \includegraphics[width=0.99\linewidth]{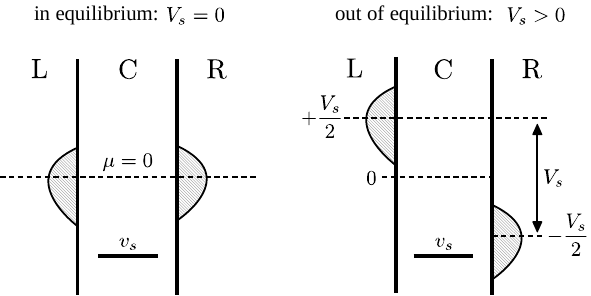}
  \caption{
    \label{fig:ks-system}
    Schematic energy diagram of effective KS potentials $v_s$ and $V_s$
    and electrode band structures (grey areas) for the nanoscale junction
    in equilibrium (left) and out of equilibrium (right)
    with symmetric voltage drop, $\pm{V_s/2}$.
  }
\end{figure}

Fig.~\ref{fig:ks-system} shows a schematic energy diagram of the KS
potentials of the system both in equilibrium (left panel) when
$V_\alpha^s=V_s=V=0$, and out of equilibrium (right panel) for a bias
voltage $V>0$ applied symmetrically between both electrodes,
$V_\LL^s=V_s/2=-V_\RR^s>0$. Note that the KS bias $V_s$ does not only shift
the chemical potentials of the leads by $\pm{V}_s/2$ but also the band
structure of the leads.
%due to electrostatic effects within the junction \textcolor{red}{[REFERENCES?]}.
Hence for a KS bias $V_s$ the embedding self-energies of the electrodes
(\ref{eq:SE_embed}) and accordingly the coupling matrices are to be
evaluated at $\omega=\pm{V_s}/2$, i.e.
$\bm\Sigma_\alpha=\bm\Sigma_\alpha(\omega+s_\alpha{V_s}/2)$ and
$\bm\Gamma_\alpha=\bm\Gamma_\alpha(\omega+s_\alpha{V_s}/2)$.  

%\subsection{Kohn-Sham equations for arbitrary bias voltage drops}

\begin{figure}
  \begin{center}
    \includegraphics[width=\linewidth]{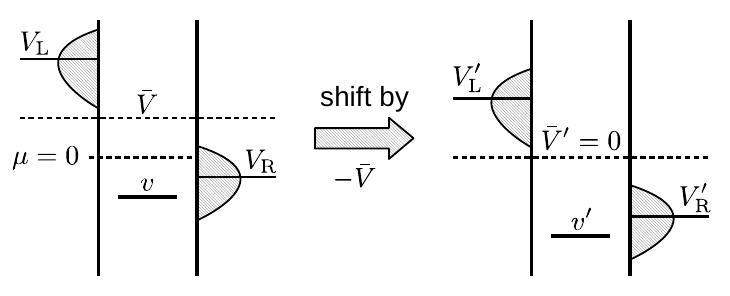}
  \end{center}
  \caption{
    \label{fig:bias_trafo}
    Transformation from a general asymmetric voltage drop $V_\LL\neq-V_\RR$ (left)
    to a symmetric voltage drop $V_\LL=V/2$ and $V_\RR=-V/2$ (right) by a total energy
    shift of $-\bar{V}=-(V_\LL+V_\RR)/2$.
  }
\end{figure}

Let us now consider the situation of an arbitrary voltage drop
across the junction, as depicted schematically in the left panel of
Fig.~\ref{fig:bias_trafo}, when the electrochemical potentials
of the two leads are not symmetrically displaced from equilibrium,
i.e. $V_\LL\neq-V_\RR$.
A transformation from an arbitrary voltage drop to a symmetric voltage
drop $V_\LL^\prime=-V_\RR^\prime=V/2$ that leaves the physical observables
unchanged can be achieved by applying a spatially constant energy shift
of the entire system by $-\bar{V}=-(V_\LL+V_\RR)/2$
relative to the equilibrium chemical potential, as depicted in Fig.~\ref{fig:bias_trafo}.
Hence the KS equations for the particle density $n(\bm{r})$ and current
$I$ for an arbitrarily distributed voltage drop $V=V_\LL-V_\RR$ can be
obtained from the KS equations for symmetric bias
(\ref{eq:dens},\ref{eq:curr}) by shifting the KS gate $v_s$ by $-\bar{V}$:
\begin{subequations}
\begin{eqnarray}
  \lefteqn{n_{\rm asym }[v,V_\LL,V_\RR](\bm{r}) =
    n_{\rm sym}[v-\bar{V},V](\bm{r}) } \nonumber\\
  &&\hspace{1ex} = 2\sum_{\alpha=\LL,\RR}\int\frac{d\omega}{2\pi}
  f\left(\omega+s_\alpha\frac{V_s}{2}\right) A_{\alpha;v_s-\bar{V}}(\bm{r};\omega) %%\nonumber\\
  \\
  \lefteqn{I_{\rm asym}[v,V_\LL,V_\RR] = I_{\rm sym}[v-\bar{V},V] } \nonumber\\ 
  &&\hspace{1ex}=  2\sum_{\alpha=\LL,\RR} s_\alpha\int\frac{d\omega}{2\pi} f\left(\omega+s_\alpha\frac{V_s}{2}\right) T_{v_s-\bar{V}}(\omega)
\end{eqnarray}
\end{subequations}
Note that the embedding self-energies and coupling matrices are still to be
evaluated at $\pm{V_s}/2$; only the KS gate $v_s$ is shifted by $-\bar{V}$.
Hence the KS GF in the transformed system is given by:
\begin{equation}
  \bm{G}_{v_s-\bar{V}}(\omega)=\frac{1}{\omega-\bm{h}_s+\bar{V}-\sum_{\alpha=\LL,\RR}\bm\Sigma_\alpha\left(\omega+s_\alpha\frac{V_s}{2}\right)}
\end{equation}

One can also introduce a KS voltage drop $V_\alpha^s$ distributed
with the same ratio as the actual voltage $V$ between both leads,
$V_\alpha^s=V_s(V_\alpha/V)$ by  performing a simple change of the
integration variable according to $\omega\rightarrow\omega-\bar{V}_s$
where $\bar{V}_s\equiv(V_\LL^s+V_\RR^s)/2$. Similarly, the arguments to the
electrode quantities (embedding
self-energies, coupling matrices and Fermi functions), change according to
$\omega-s_\alpha{V_s/2}\rightarrow\omega-V^s_\alpha$.
We thus obtain the i-DFT KS equations for asymmetric voltage drops:

\begin{subequations}
  \begin{eqnarray}
    n_{\rm asym }[v,V_\LL,V_\RR](\bm{r}) &=& 2\sum_{\alpha}\int\frac{d\omega}{2\pi}
    f\left(\omega-V_\alpha^s\right) \bar{A}^\alpha_{v_s-\bar{V}_\xc}(\bm{r};\omega) \nonumber\\
    \label{eq:dens_as}\\
    I_{\rm asym }[v,V_\LL,V_\RR] &=&  2\sum_{\alpha} s_\alpha \int\frac{d\omega}{2\pi} f\left(\omega-V_\alpha^s\right) \bar{T}_{v_s-\bar{V}_\xc}(\omega) \nonumber\\
    \label{eq:curr_as}
  \end{eqnarray}
\end{subequations}
where $\bar{V}_\xc=\bar{V}_s-\bar{V}$ is the average xc bias of both leads
and the spectral function $\bar{A}^\alpha_{v_s}$ and transmission function $\bar{T}_{v_s}$
refer to a new KS GF for arbitrary chemical potentials $V^s_\alpha$ of the leads
\begin{equation}
  \bm{\bar{G}}_{v_s}(\omega)=\frac{1}{\omega-\bm{h}_s-
    \sum_{\alpha=\LL,\RR}\bm\Sigma_\alpha\left(\omega-V_\alpha^s\right)}.
\end{equation}

\subsection{i-DFT functionals for the Constant Interaction Model at asymmetric coupling}

We now apply the i-DFT formalism described above 
to the so-called constant interaction model (CIM) which is widely used as 
a model for the description of effects of strong electronic correlation such 
as Coulomb blockade (CB) or the Kondo effect. 
The CIM Hamiltonian can be obtained from Eq.~(\ref{eq:H_C}) by simplifying 
the Coulomb interaction in the central region  according to
$U_{ijkl}=U\delta_{ik}\delta_{jl}$. Additionally we assume a diagonal
one-body part $h^0_{ij}=\varepsilon_i\delta_{ij}$ which leads to
\begin{equation}
 \label{eq:CIM}
 \mathcal{H}_\CC^{\rm CIM} = \sum_{i\sigma} \varepsilon_i \, \hat{n}_{i\sigma} + \frac{U}{2} \sum_{i\sigma\neq{j\sigma^\prime}} \hat{n}_{i\sigma} \hat{n}_{j\sigma^\prime}
\end{equation}
where $\hat{n}_{i\sigma}$ is the electron occupation operator 
for level $i$ with spin $\sigma$. We furthemore assume
that the coupling matrices $\bm\Gamma_{\alpha}$ are energy independent,
i.e., we are in the wide-band limit (WBL) for both leads. Note that for a 
single level this becomes the single-impurity Anderson model 
(SIAM)\cite{Anderson:PR:1961}. 

\begin{figure}
  \includegraphics[width=0.99\linewidth]{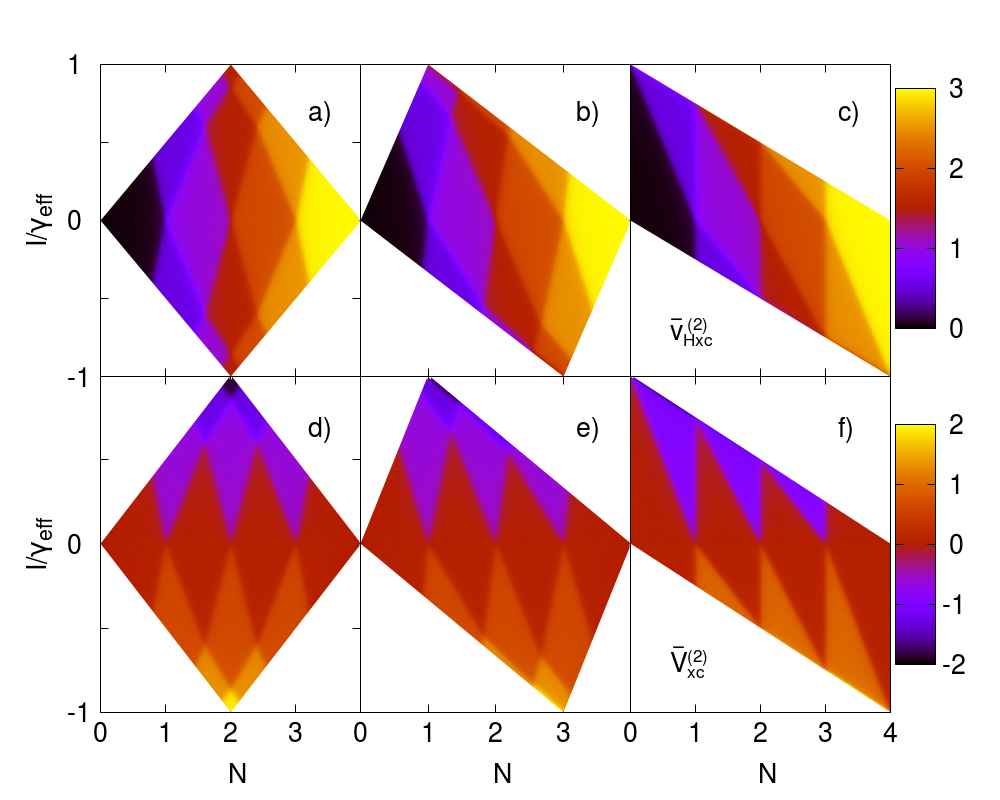}
  \caption{Model Hxc gate (upper panels) and xc bias potentials (lower panels) 
    for a two-level CIM ($\mathcal{M}=2$) according to 
    Eqs.~(\protect\ref{xcgate_CIM}) and 
    (\protect\ref{xcbias_CIM}), respectively, for different asymmetric 
    coupling at fixed $\gamma=\gamma_{\rm L}+\gamma_{\rm R}$ and $U/\gamma=5$. 
    Panels a) and d): $\gamma_{\rm L}/\gamma=0.5$. Panels b) and e): 
    $\gamma_{\rm L}/\gamma=0.25$. Panels c) and f): 
    $\gamma_{\rm L}/\gamma=5 \times 10^{-5}$. Current in units of 
    $\gamma_{\rm eff}=4 \gamma_{\rm L}\gamma_{\rm R}/\gamma$, xc potentials 
    in units of $U$. }
  \label{fig:xcpots_asymm}
\end{figure}

The CIM has been studied within the i-DFT framework both 
in the Coulomb blockade \cite{StefanucciKurth:15} as well as in the Kondo 
regime \cite{Kurth:PRB:2016,Kurth:JPCM:2017} and approximate i-DFT 
xc potentials have been suggested. However, these approximations 
were restricted to symmetric coupling. In a very recent work
\cite{Jacob:NL:2018} we have designed an approximation for the
limiting case of an extremely asymmetrically coupled CIM where the coupling
to one of the leads becomes infinitesimal. We were interested in this extreme
limit because it can be shown that in this limit one can extract equilibrium
many-body spectral functions at zero temperature from differential
conductances computed with i-DFT \cite{Jacob:NL:2018}. In the present work we
are interested in arbitrary asymmetry in the coupling and we thus 
need to construct xc potentials for this case as well. 
For simplicity, we restrict ourselves to the CIM with an arbitrary number 
of degenerate single-particle levels ($\varepsilon_i=\varepsilon$) which are 
all coupled in the same way to the lead $\alpha$, i.e., the coupling 
matrices in the single-particle basis 
$\bm\Gamma_{\alpha} = \gamma_\alpha {\mathbf 1}$ are proportional to the unit 
matrix $\mathbf 1$ (but the constants $\gamma_{\rm L}$
and $\gamma_{\rm R}$ may differ).
In this case the i-DFT xc potentials depend only on the total number 
$N=\sum_{i\sigma} n_{i\sigma}$ of electrons on the dot. In 
Ref.~\cite{StefanucciKurth:15} we constructed xc functionals for the 
Coulomb blockade regime by numerical inversion of rate equations 
\cite{Beenakker:PRB:1991}. The resulting xc potentials showed a complex pattern of 
smeared steps of height $U/2$ for the Hxc gate and height $U$ for the xc 
bias potential. The resulting parametrization for symmetric coupling can also 
be used for asymmetric coupling with appropriate modifications described 
below. For a dot with $\mathcal{M}$ levels this parametrization reads 
\begin{subequations}
  \begin{eqnarray}
    \bar{v}^{(\mathcal{M})}_{\rm Hxc}[N,I]&=&\frac{U}{4}\sum_{K=1}^{2\mathcal{M}-1}
    \sum_{s=\pm}\!\left[1+\frac{2}{\pi}\,
      {\rm atan}\frac{\Delta_{K}^{(s)}(N,I)}{W}\right]\quad\quad
    \label{xcgate_CIM}
    \\
    \bar{V}^{(\mathcal{M})}_{\rm 
      xc}[N,I]&=&-U\sum_{K=1}^{2\mathcal{M}-1}\sum_{s=\pm}\frac{s}{\pi}\,
        {\rm atan}\frac{\Delta_{K}^{(s)}(N,I)}{W}\quad\quad
        \label{xcbias_CIM}
  \end{eqnarray}
  \label{xc_CIM}
\end{subequations}
with the fit parameter $W=0.16 \gamma/U$ with
$\gamma = \gamma_{\rm L} + \gamma_{\rm R}$.
The $\Delta_{K}^{(s)}(N,I)$ are piecewise linear functions 
of $N$ and $I$ of positive ($s=+$) and negative ($s=-$) slopes connecting 
vertices in the $N$-$I$ plane and passing through the vertex at $(K,0)$ with 
$K$ integer. For symmetric coupling, explicit expressions for the vertices 
could be given \cite{StefanucciKurth:15}. For asymmetric coupling, we 
determine the vertices by solving the rate equations \cite{Beenakker:PRB:1991}
for its density-current plateau values. These plateau values result when the 
Fermi functions with argument $\varepsilon + (K-1)U \pm V/2$, 
$K \in [1,2\mathcal{M}]$ entering the rate equations either vanish or are 
equal to unity. Taking into account the consistency condition 
that if the Fermi factor for a given argument is unity then all other Fermi 
factors with smaller argument have to take the same value then 
one correctly obtains $(2\mathcal{M}+1)^2$ vertices in the $N$-$I$ plane. In 
Fig.~\ref{fig:xcpots_asymm} we show the evolution of the (H)xc gate and bias 
potentials obtained in this way for a degenerate two-level CIM
($\mathcal{M}=2$) with increasing asymmetry in the coupling. 
First we note that the codomain for allowed $(N,I)$ values is deformed: the 
maximum current of value $\mathcal{M} \gamma_{\rm eff}/2$ with 
$\gamma_{\rm eff}=4 \gamma_{\rm L}\gamma_{\rm R}/\gamma$ occurs at density 
$\mathcal{M} \gamma_{\rm L}/\gamma$, the minimum current $-\mathcal{M} 
\gamma_{\rm eff}/2$ at density $\mathcal{M} \gamma_{\rm R}/\gamma$. With 
increasing asymmetry, the vertices with {\em finite} current move more and 
more towards the edges of the codomain, leading to a relatively simple pattern 
of steps in the extreme limit (panels c) and f)) such that the corresponding 
$\Delta_{K}^{(s)}(N,I)$ can be given analytically~\cite{Jacob:NL:2018}. 

The model (H)xc potentials of Eq.~(\ref{xc_CIM}) are constructed by reverse
engineering of rate equations. Therefore they contain Coulomb blockade but 
no Kondo physics. In a DFT framework, the Kondo effect in the zero-bias 
conductance of weakly coupled quantum dots is already captured correctly in 
the KS conductance, both for single-level 
\cite{Stefanucci:PRL:2011,Bergfield:PRL:2012,Troester:PRB:2012} as well as
for multi-level dots \cite{StefanucciKurth:13}. The incorporation of Kondo
physics into the i-DFT functionals thus requires that the derivative of the
xc bias w.r.t. the current at $I=0$ vanishes \cite{StefanucciKurth:15}.
In Ref.~\cite{Kurth:PRB:2016} we proposed modified (H)xc potentials to include
Kondo physics. A further straightforward generalization of these functionals
to asymmetric coupling then takes the form (at zero temperature)
\begin{subequations}
\begin{equation}
v^{(\mathcal{M})}_{\rm Hxc,1}[N,I] = (1 - a_1[I]) \bar{v}^{(\mathcal{M})}_{\rm Hxc}[N,I] 
+ a_1[I] \bar{v}_{\rm Hxc}[N]
\label{eq:xcgate_CIM_mod1}
\end{equation}
\begin{equation}
V^{(\mathcal{M})}_{\rm xc,1}[N,I] = (1 - a_1[I]) \bar{V}^{(\mathcal{M})}_{\rm xc}[N,I] \;.
\label{eq:xcbias_CIM_mod1}
\end{equation}
\label{xc_CIM_mod1}
\end{subequations}
with the the purely current-dependent function
\begin{equation}
  a_1[I] = 1 - \left[\frac{2}{\pi} \arctan\left(
    \frac{I}{W  \gamma_{\rm eff}} \right) \right]^2
\label{afunc1}
\end{equation}
In Eq.~(\ref{eq:xcgate_CIM_mod1}), the zero-current Hxc potential 
$\bar{v}_{\rm Hxc}$ is defined as 
$\bar{v}_{\rm Hxc}[N] = \sum_{K=1}^{2{\mathcal{M}-1}} v_{\rm Hxc}^{\rm ext}[N-(K-1)]$ 
where the extended function is 
\begin{equation}
v_{\rm Hxc}^{\rm ext}[N] =
\left\{
\begin{array}{cl}
0 & \mbox{ $N<0$} \\
v_{\rm Hxc}^{\rm SIAM}[N] & \mbox{ $0\leq N \leq 2$}~~, \\
U & \mbox{ $N>2$}
\end{array}
\right.
\end{equation}
and $v_{\rm Hxc}^{\rm SIAM}[N]$ is the parametrization of the equilibrium SIAM
Hxc potential of Ref.~\cite{Bergfield:PRL:2012}. As a final tweak, we have
made the substitution $W\to 2W$ in both $\bar{v}^{(\mathcal{M})}_{\rm Hxc}[N,I]$ and 
$\bar{V}^{(\mathcal{M})}_{\rm xc}[N,I]$ used in Eqs.~(\ref{xc_CIM_mod1}).
This latter modification has been introduced in Ref.~\cite{Kurth:PRB:2016} in
order to better reproduce accurate differential conductances of the SIAM at
symmetric coupling. 

In Ref.~\cite{Jacob:NL:2018} we derived the following exact condition
\begin{equation}
  \label{eq:condition}
  \lim_{\Gamma_{\rm L}\rightarrow0} v_{\rm Hxc}[n,I](\bm{r}) + 
  \frac{1}{2}V_{\rm xc}[n,I] = v_{\rm Hxc}^{(0)}[n](\bm{r}) \;.
\end{equation}
which relates the (current-dependent) xc gate and bias potentials in the
extremely asymmetric limit to the Hxc gate potential in the ground state.
The reason we are interested in this particular limit is based on the fact
that at zero temperature in this limit and with the bias completely applied
to the weakly coupled lead, one can relate the differential conductance to
the equilibrium spectral function of the nanoscale region C through
\cite{Jacob:NL:2018}
\begin{equation}
  A(\omega) = \lim_{\gamma_{\rm L} \to 0} \frac{ 4 \pi}{\gamma_{\rm eff}}
  \frac{\partial{I}}{\partial{V}} \bigg\vert_{V=\omega}
  \label{eq:specfunc}
\end{equation}
where $A(\omega)$ is the trace of the {\em many-body} spectral function
matrix. Note that $\gamma_{\rm eff}\rightarrow4\,\gamma_\LL$ for $\gamma_\LL\rightarrow0$.
Computing the differential conductance from i-DFT thus allows to extract
the many-body spectral function from a DFT framework. 

Unfortunately, our functionals of Eq.~(\ref{xc_CIM_mod1}) do not satisfy
this condition (although the ones of Eq.~(\ref{xc_CIM}) do). Therefore here
we propose an alternative but similar functional for which this condition
holds by construction, i.e., 
\begin{subequations}
\begin{eqnarray}
v^{(\mathcal{M})}_{\rm Hxc,2}[N,I] &=& (1 - a_2[I]) \left( 
\bar{v}^{(\mathcal{M})}_{\rm Hxc}[N,I] - \bar{v}^{(\mathcal{M})}_{\rm Hxc}[N,0] \right) 
\nonumber \\
&& + \bar{v}_{\rm Hxc}[N]
\label{eq:xcgate_CIM_mod}
\end{eqnarray}
\begin{equation}
V^{(\mathcal{M})}_{\rm xc,2}[N,I] = (1 - a_2[I]) \bar{V}^{(\mathcal{M})}_{\rm xc}[N,I] \;.
\label{eq:xcbias_CIM_mod}
\end{equation}
\label{xc_CIM_mod}
\end{subequations}
In principle, in Eq.~(\ref{xc_CIM_mod}) we could have used the same function
$a_1[I]$ (Eq.~(\ref{afunc1})) as used in the xc potentials of
Eq.~(\ref{xc_CIM_mod1}). However, in order to better reproduce accurate
equilibrium spectral functions for the SIAM (in the extremely asymmetric
limit), we were compelled to use the alternative function $a_2[I]$ as 
\begin{equation}
  a_2[I] = 1 - \frac{2}{\pi} \arctan\left[
    \lambda \left(\frac{I}{W  \gamma_{\rm eff}} \right)^2 \right]
\label{afunc2}
\end{equation}
where $\lambda=0.16$ is a fit parameter. 

\section{Results}

\begin{figure}[t]
  \begin{center}
    \includegraphics[width=0.99\linewidth]{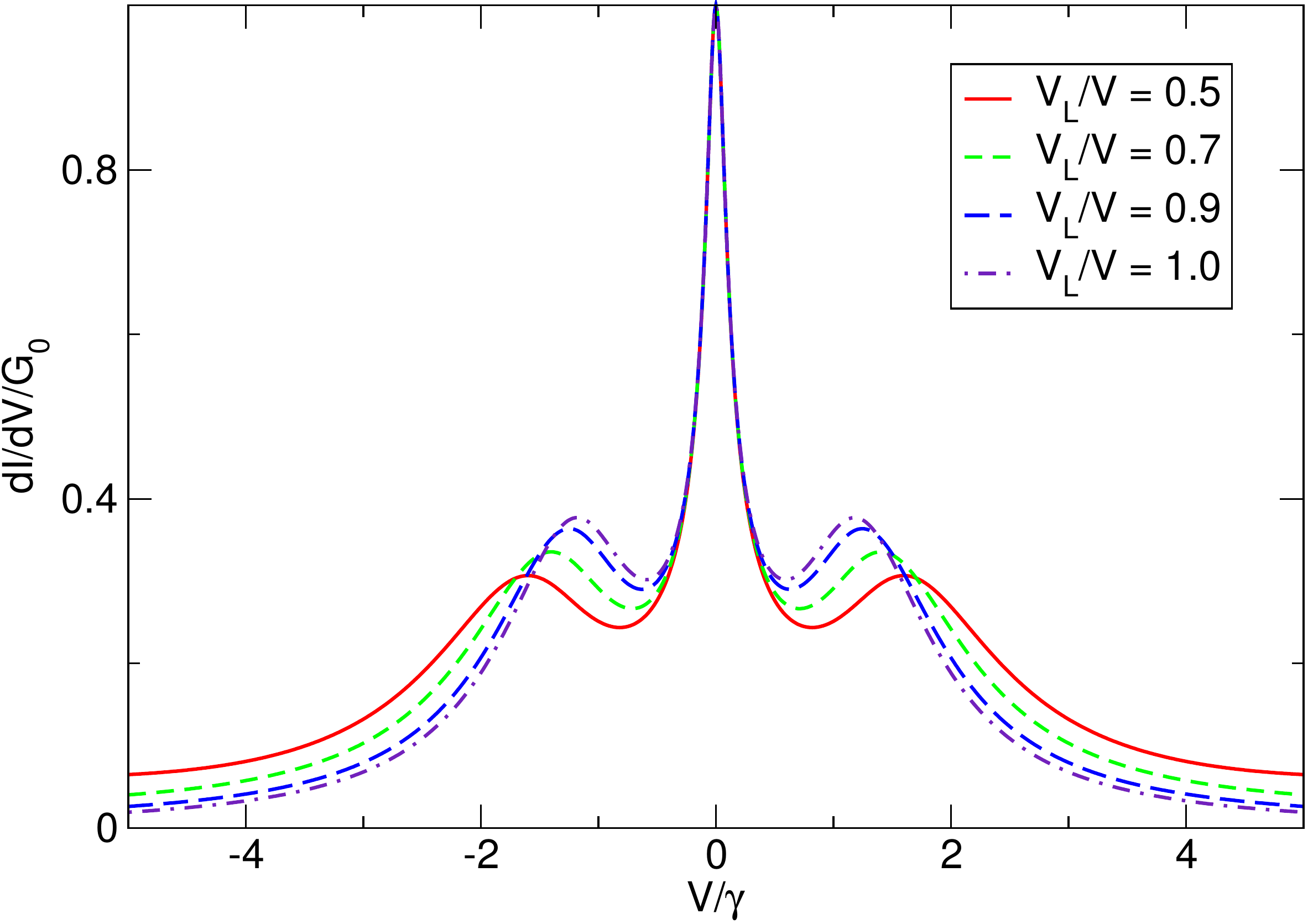}
  \end{center}
  \caption{
    Evolution of the zero-temperature differential conductance of i-DFT using
    the functional of Eq.~(\ref{xc_CIM_mod}) for the 
    symmetrically coupled SIAM with $U/\gamma=3$ at particle-hole symmetry 
    from symmetric to asymmetric voltage drops using the functional
    of Eq.~(\ref{xc_CIM_mod}).
    %%Note that for better visibility the spectra an
    %%increasing offset has been added.
  }
  \label{fig:bias_asymm}
\end{figure}

In the present section we will show some results obtained with the
functionals described above. Our main focus is on the SIAM which has been
studied with many different methods and therefore we can compare the
results of our i-DFT approach with accurate reference calculations. 

We start with results for the SIAM. As a first example, in
Fig.~\ref{fig:bias_asymm} we show how the zero-temperature differential
conductance of the symmetrically coupled SIAM at fixed gate $\varepsilon=-U/2$
(particle-hole symmetry) evolves as the bias asymmetry between left and right
lead increases. The results were obtained with i-DFT using the functional of
Eq.~(\ref{xc_CIM_mod}) for interaction strength $U/\gamma=3$. As the asymmetry
in the applied bias increases, the side peaks move to lower biases. The Kondo
resonance around $V=0$, on the other hand, is not affected by the bias
asymmetry.
%%{\color{red}Maybe this last point is easier to see without the
%%  offset...?}

\begin{figure}[t]
  \begin{center}
    \includegraphics[width=0.99\linewidth]{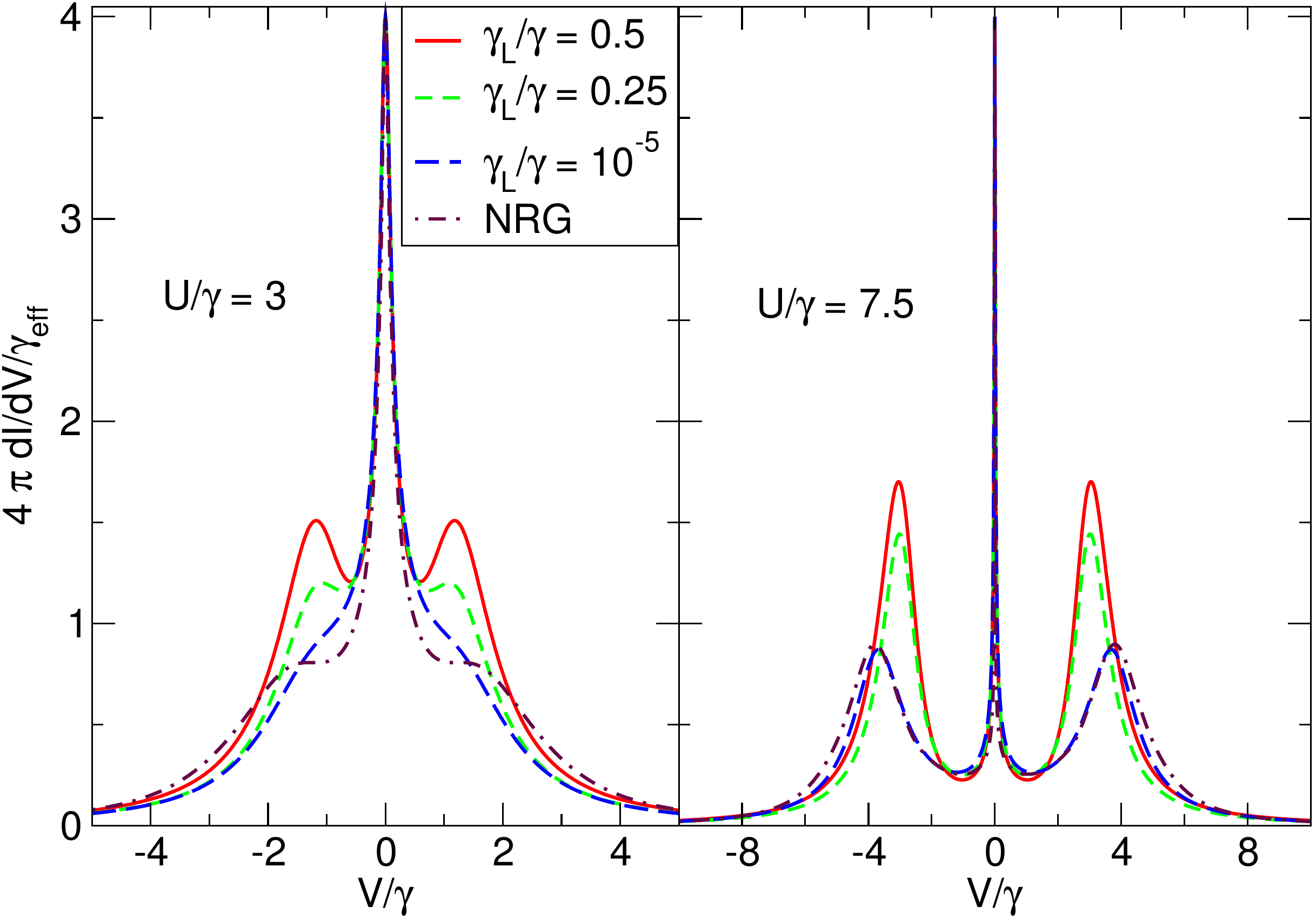}
  \end{center}
  \caption{
    Evolution of the zero-temperature differential conductance of the SIAM
    from symmetric to completely asymmetric coupling at completely asymmetric
    voltage drop ($V_\LL=V$) for different interaction strengths $U/\gamma$.
    In both cases, the external gate is fixed at particle-hole symmetry,
    $\varepsilon=-U/2$, and the functional of Eq.~(\ref{xc_CIM_mod}) has been used.
    For comparison, equilibrium spectral functions from numerical
    renormalization group (NRG) calculations of Ref.~\cite{Motahari:PRB:2016}
    are shown.
  }
  \label{fig:coup_asymm}
\end{figure}

In Fig.~\ref{fig:coup_asymm} we show the zero-temperature differential
conductance of the SIAM for different asymmetry in the coupling to left and
right leads obtained with the i-DFT functional of Eq.~(\ref{xc_CIM_mod}).
Results are shown for two values of the interaction strength, both at
particle-hole symmetry and at completely asymmetric voltage drop ($V_\LL=V$).
With increasing coupling asymmetry, the maxima of the side peaks decrease
significantly and, for $U/\gamma=7.5$, are shifted slightly towards higher
biases. For comparison we also show equilibrium spectral functions obtained
with Numerical Renormalization Group (NRG) methods \cite{Motahari:PRB:2016}.
While for the lower interaction strength ($U/\gamma=3$), the i-DFT differential
conductance for the highly asymmetrically coupled case
$\gamma_{\rm L}/\gamma = 10^{-5}$ exhibits side peaks (actually more side
shoulders) which agree moderately well with the NRG spectral function, for the
higher interaction strength ($U/\gamma=7.5$), the agreement between
the i-DFT and NRG spectral functions is actually quite remarkable. 

\begin{figure}[t]
  \begin{center}
    \includegraphics[width=0.99\linewidth]{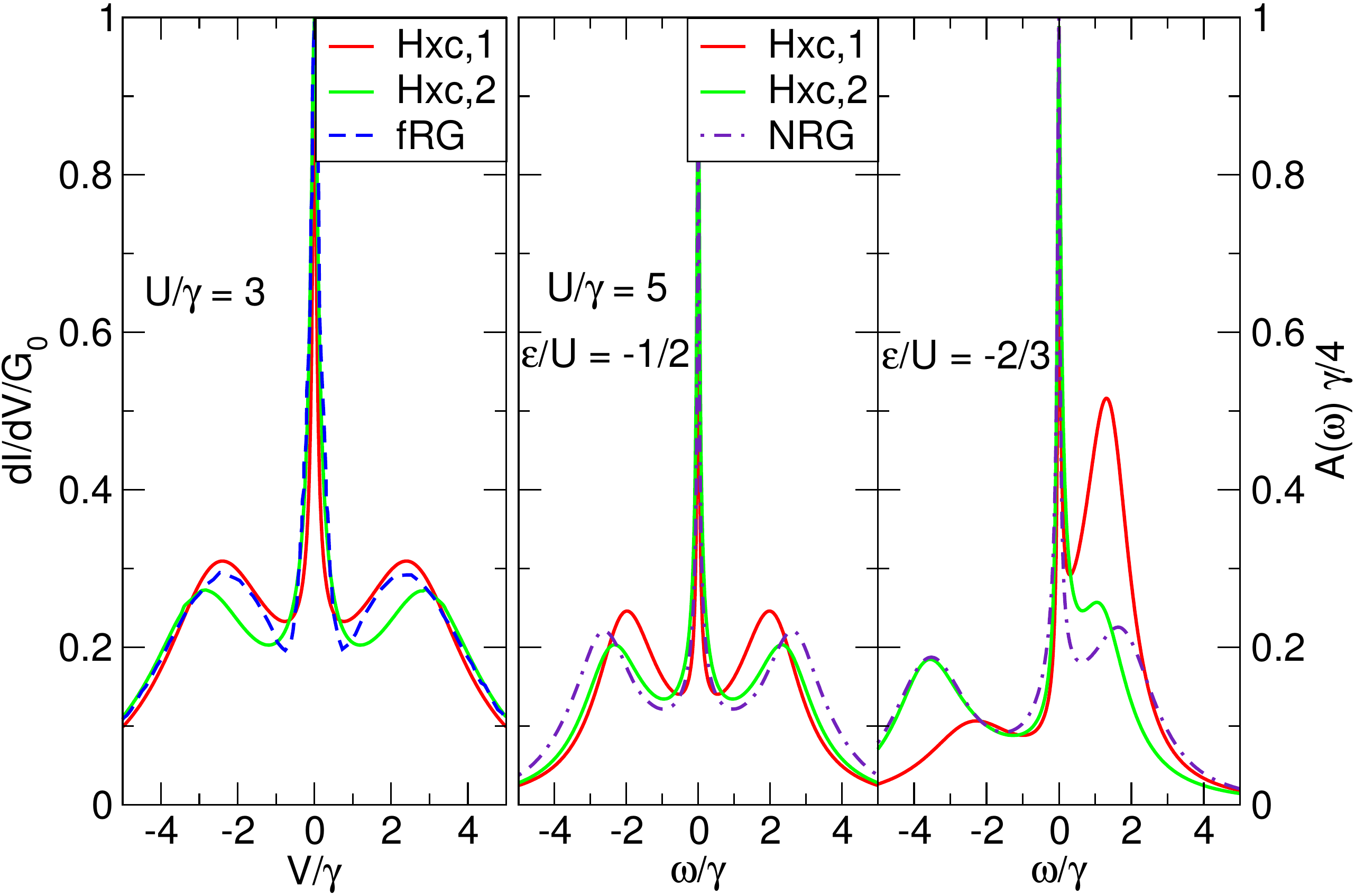}
  \end{center}
  \caption{Comparison of the performance of the two different i-DFT
    functionals of Eq.~(\ref{xc_CIM_mod1}) and (\ref{xc_CIM_mod})
    (denoted as ``Hxc,1'' and ``Hxc,2'', respectively).
    Left panel: differential conductance of the symmetrically coupled and
    symmetrically biased SIAM for $U/\gamma=3$ compared to results obtained
    with the functional renormalization group (fRG)
    \cite{JakobsPletyukhovSchoeller:10}. Middle and right
    panels: spectral functions for $U/\gamma=5$ at (middle panel) and away
    from (right panel) ph symmetry. NRG results from
    Ref.~\cite{Motahari:PRB:2016} for comparison. 
  }
  \label{fig:comp_KS16_JK18}
\end{figure}

In Fig.~\ref{fig:comp_KS16_JK18} we compare the performance of the i-DFT
functionals of Eq.~(\ref{xc_CIM_mod1}) and (\ref{xc_CIM_mod}). In the left
panel we compare differential conductances at symmetric coupling and bias for
$U/\gamma=3$ at particle-hole symmetry and compare with accurate results from
the functional Renormalization Group (fRG) \cite{JakobsPletyukhovSchoeller:10}.
In the other two panels we compare i-DFT spectral functions (i.e., asymmetric
coupling and bias) for $U/\gamma=5$
at (middle panel) and away (right panel) from particle-hole symmetry with
NRG results \cite{Motahari:PRB:2016}. While for the differential conductance in
the symmetrically coupled case, the functional of Eq.~(\ref{xc_CIM_mod1})
performs somewhat better than the one of Eq.~(\ref{xc_CIM_mod}), in the case
of the spectral functions the situation is just the opposite. In particular,
away from particle-hole symmetry (right panel), the functional
(\ref{xc_CIM_mod1}) strongly overestimates the side peak for positive frequency
while at the same time underestimating the one at negative frequency. Both
i-DFT functionals have the tendency to shift the side peaks to lower
frequencies. The fact that one of the functionals (Eq.~(\ref{xc_CIM_mod1}))
performs better for symmetric coupling while the other one
(Eq.~(\ref{xc_CIM_mod})) performs better for spectral functions
(highly asymmetric coupling and bias) is not really surprising since the
corresponding functions $a_1[I]$ and $a_2[I]$ (Eqs.~(\ref{afunc1}) and
(\ref{afunc2}), respectively) were actually chosen to perform well in
exactly the situation where they do.
Of course, it would be desirable to
have {\em one} functional which performs best for all the different situations
but so far this has turned out to be elusive. Nevertheless, it is still
significant progress that one can obtain from a DFT framework reasonable
differential conductances and even (equilibrium) spectral functions
for a strongly correlated system like the SIAM.

\begin{figure}[t]
  \begin{center}
    \includegraphics[width=0.99\linewidth]{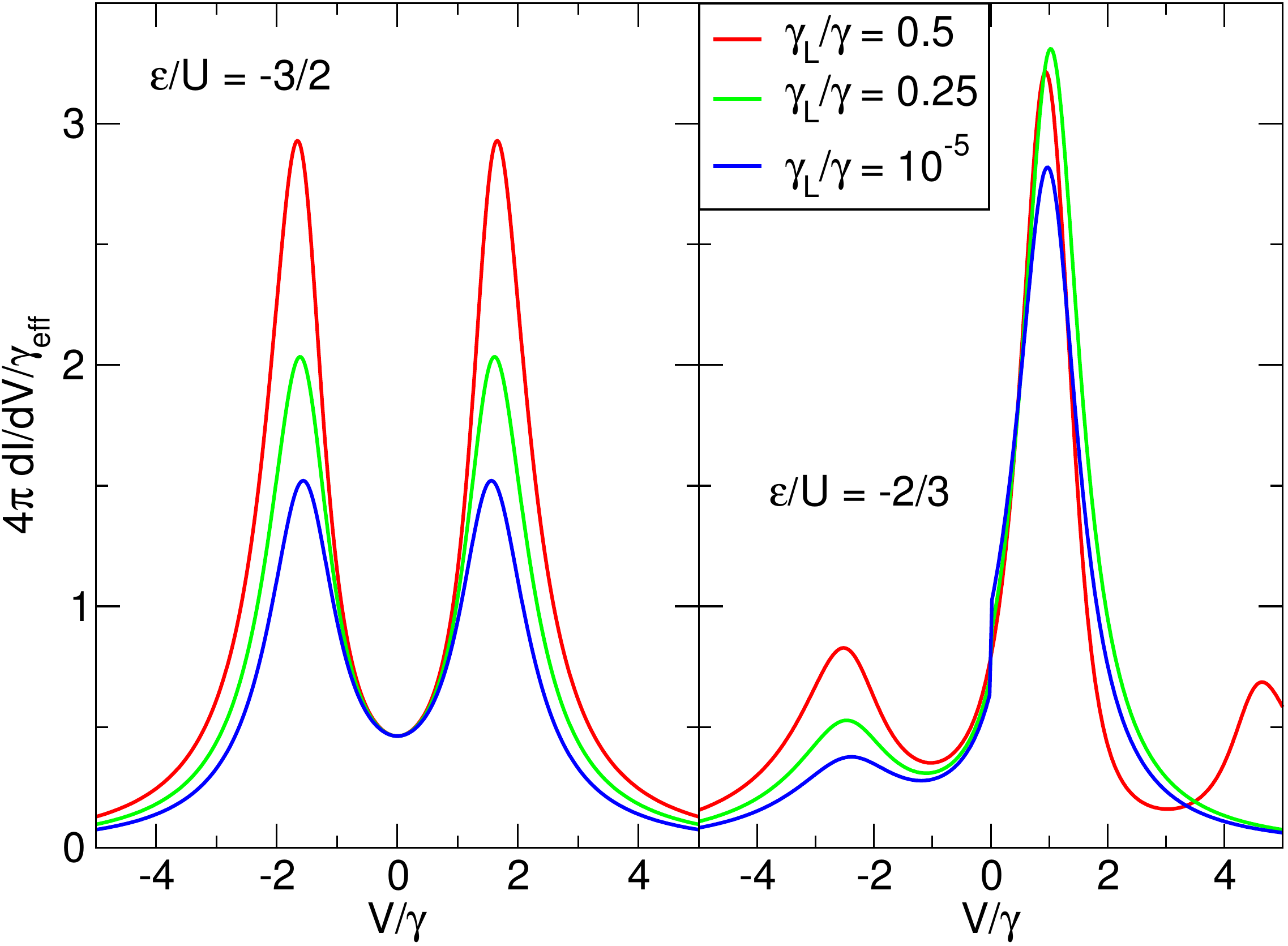}
  \end{center}
  \caption{Differential conductance for a degenerate two-level CIM with
    $U/\gamma=3$ for completely asymmetric bias and different asymmetric
    couplings obtained with the functional of Eq.~(\ref{xc_CIM}) 
    at (left panel) and away from particle-hole symmetry (right panel). 
  }
  \label{fig:2_level_CIM_CB}
\end{figure}

In Fig.~\ref{fig:2_level_CIM_CB} we show i-DFT results for differential
conductances (normalized by a factor $4\pi/\gamma_{\rm eff}$ of a CIM with
$\mathcal{M}=2$ degenerate single-particle levels at completely
asymmetrically applied bias and a various asymmetric couplings.
Here we have used the functional of Eq.~(\ref{xc_CIM}) (see also
Fig.~\ref{fig:xcpots_asymm}) which includes Coulomb blockade but no Kondo
physics. At particle-hole symmetry (left panel) the normalized differential
conductances for different coupling asymmetries are qualitatively similar
although the height of the Coulomb blockade sidepeaks decreases with
increasing coupling asymmetry.
On the other hand, the normalized differential conductance
  at zero bias is independent of the coupling asymmetry, since
  it is determined by the equilibrium spectral density at the Fermi level,
  which only depends on the total broadening $\gamma_\LL+\gamma_\RR$. 
%% At bias $V=0$ the (normalized) differential
%% conductance is independent of the coupling asymmetry since in this case we
%% have density $N=2$ and
%% $\frac{\partial \bar{V}_{\rm xc}^{\mathcal(2)}}{\partial I}\big\vert_{I=0}=0$.
%% Since the total zero-bias conductance is \cite{StefanucciKurth:15}
%% \begin{equation}
%%   G = \frac{G_s}{1 - G_s
%%     \frac{\partial \bar{V}_{\rm xc}^{\mathcal(2)}}{\partial I}\big\vert_{I=0}}
%% \end{equation}
%% where $G_s$ is the KS zero-bias conductance this means that at $N=2$ and in
%% the zero-bias limit $G=G_s$ which is solely determined by the
%% {\em equilibrium} density and thus independent of the coupling asymmetry.
%% {\color{red}Does this argument make sense?}
Away from particle-hole symmetry (right panel) the normalized differential
conductances also qualitatively changes with increasing coupling asymmetry:
while for symmetric coupling ($\gamma_{\rm L}/\gamma=0.5$) we have a three-peak
structure in the $\partial{I}/\partial{V}$, in the highly asymmetric case
($\gamma_{\rm L}/\gamma=10^{-5}$) this has been transformed into a two-peak
structure. The latter fact is not surprising: for the highly asymmetric limit
the density is essentially independent of the bias and, for fixed density, the 
corresponding xc bias exhibits exactly two steps as the current is varied 
(see panel f) of Fig.~\ref{fig:xcpots_asymm}). 
These steps are the origin of the peaks in the differential conductance.

It is worth pointing out that the apparent discontinuities in the differential
conductances at zero bias and away from particle-hole symmetry are an artefact
of our parametrization, Eq.~(\ref{xc_CIM}), where the
$\Delta^{(\pm)}_{K}(N,I)$ are approximated as piecewise linear functions of
$N$ and $I$. As one
can appreciate in Fig.~\ref{fig:xcpots_asymm}, these lines (the positions
of the steps in the Hxc gate and xc bias) in most cases have kinks when
crossing one of the vertices in the $N$-$I$ plane. These kinks are the
origin of the discontinuities in the $\partial{I}/\partial{V}$ curves.
Therefore, if one parametrized the $\Delta^{(\pm)}_{K}(N,I)$ as
differentiable functions of $N$ and $I$ these discontinuities would disappear.

%% {\color{red} Should we mention the problems with our other functionals or
%%   should we wait for the referee to spot this omission?}

\section{Conclusions}

The recently proposed i-DFT approach provides a promising framework for the 
DFT description of steady-state transport for both weakly and strongly 
correlated systems. As usual for any DFT, the crucial ingredient for a 
successful application is the quality of the approximations for the 
exchange-correlation functionals. In previous work, functionals have been
constructed for the Anderson model and the Constant Interaction Model for
the situation when the coupling to the leads is either symmetric or for the
limiting case when one of the leads is extremely weakly coupled. In the
present work we have generalized these functionals to arbitrary asymmetry and
compared their relative performance. This has been achieved
by first constructing a functional which is able to capture Coulomb
blockade at arbitrary coupling and bias using insights gained from
reverse engineering of the rate equation approach. As a second step, Kondo
physics can be introduced into the i-DFT functionals in a relatively easy
manner by the requirement that the derivative of the xc bias with respect to
the current vanishes in the zero-current limit which we have done in two different
ways. 

In the present as well as in previous work, the i-DFT formalism has been
applied to (minimal) model systems describing transport through correlated
systems. However, typically one thinks of DFT as a method for an atomistic
description of molecular or solid state systems. It is therefore one of the
pending tasks of i-DFT to translate the insights gained for model systems
into workable approximations which can be applied to an atomistic description
of electronic transport.

\section*{Acknowledgements}
We acknowledge funding through the grant
``Grupos Consolidados UPV/EHU del Gobierno Vasco'' (IT578-13).
S.K. additionally acknowledges funding through a grant of the
"Ministerio de Economia y Competividad (MINECO)" (FIS2016-79464-P).

\bibliographystyle{phaip}
\bibliography{nanodmft,nano_bibfile}

\end{document}